\documentclass[10pt, pre, amsmath, onecolumn, showpacs]{revtex4-1}

\usepackage{graphicx}
\usepackage{amsmath,amsfonts,amssymb,color}

\usepackage{hyperref}
\usepackage{footnote}
\usepackage{graphicx}
\usepackage{dcolumn}
\usepackage{bm}
\usepackage[caption=false]{subfig}

\begin{document}


\title{Fluctuation-dominated phase ordering at a mixed order transition}

\author{Mustansir Barma$^1$, Satya N. Majumdar$^2$ and David Mukamel$^3$ }

\affiliation{$^1$ TIFR Centre for Interdisciplinary Sciences, Tata 
Institute of Fundamental Research, Gopanpally, Hyderabad 500107, India\\
$^2$ LPTMS, CNRS, Univ. Paris-Sud,
Universit\'{e} Paris-Saclay, 91405 Orsay, France \\
$^3$ Department of Physics of Complex Systems, 
Weizmann Institute of Science, Rehovot 7610001, Israel}

\begin{abstract}
	
Mixed order transitions are those which show a discontinuity of the order 
parameter as well as a divergent correlation length. We show that the 
behaviour of the order parameter correlation function
along the transition line of mixed order transitions can change from normal critical behaviour with power law decay, to fluctuation-dominated phase ordering as a parameter 
is varied. The  defining features of fluctuation-dominated order are anomalous 
fluctuations which remain large in the thermodynamic limit, and 
correlation functions which approach a finite value through a cusp 
singularity as the separation scaled by the system size approaches zero. 
We demonstrate that fluctuation-dominated order sets in along a portion of 
the transition line of an Ising model with truncated long-range 
interactions which was earlier shown to exhibit mixed order transitions, 
and also argue that this connection should hold more generally.

\end{abstract}
\maketitle

\section{Introduction}

The locus which separates an ordered state from a disordered state with a diverging correlation length is usually 
characterized by power-law decays of correlation functions, indicative of critical behaviour. However, a growing 
number of systems show a related but distinct behaviour, termed {\it fluctuation-dominated phase ordering} (FDPO), 
along the critical line. (We continue to refer to the order-disorder separatrix as a critical line, even if the 
behaviour along it sometimes departs from traditional critical behaviour.) The crucial distinction is 
that in FDPO, 
the two-point correlation function of the order parameter $G(r|L)$ does not decay as a simple power, but rather is a scaling 
function of separation $r$ scaled by the system size $L$, in the limit $r \rightarrow \infty, L \rightarrow \infty$ 
with the ratio $\frac{r}{L}$ held constant; as this ratio approaches zero, the scaling function approaches a 
constant value in a singular fashion, through a cusp singularity:
\begin{equation} 
 G(r|L) \approx m_0^2 - a\, \left|\frac{r}{L}\right|^{\alpha} + \cdots
\label{Eq_Cusp}
\end{equation}
where $m_0$ is the order parameter along the critical line, defined through the asymptotic behaviour of the two-point correlation function in an infinite system.
The cusp exponent $\alpha$ lies between 0 and 1 and varies from system to system. This signature of FDPO has been 
found in several non-equilibrium models, ranging from particles on fluctuating surfaces \cite{DB,DBM} to active 
nematics \cite{MR,DDR}, granular collisions \cite{SDR} and proteins on a cell surface \cite{DPR}. A cusp in the
correlation function also arises in disordered systems such as porous  solids \cite{PZW}, rough films \cite{AB1} and
random-field Ising systems \cite{AB2}. 
The cusp singularity implies that the Porod Law ($\alpha = 1$), familiar in the study of phase ordering dynamics \cite{AJB}, does not hold; its breakdown is associated with the formation of anomalously large interfacial regions between ordered phases. The other principal 
characteristic of FDPO is the occurrence of very large fluctuations \cite{DBM,DDR}, leading to a broad distribution 
of the order parameter \cite{DBM,KBB} as well as some other observables \cite{DBM,CB}.

In this paper we explore the connection between FDPO and {\it mixed order transitions} (MOTs). These transitions are 
characterized by a discontinuity of the order parameter as in first order phase transitions together with a 
diverging correlation length as in second order transitions. Examples of mixed order transitions include some 
discrete spin models with long-range interactions \cite{DM1, DM2, DM3, DM4, DM5}, models of depinning transitions 
such as DNA denaturation \cite{PS, MF} and wetting transitions \cite{WT1, WT2}. More recent studies of glass and 
jamming transitions \cite{JT1, JT2, JT3, JT4, JT5}, evolution of complex networks \cite{CN1,CN2,CN3,CN4} and active 
polymer gels \cite{AP} have shown that mixed order transitions take place in such systems as well. While all these 
systems do exhibit MOT, they differ in some of their features. In particular two broad classes of systems have been 
observed. In one class the correlation length diverges rather sharply, with essential singularity, as the transition 
is approached while in the other the divergence is algebraic. A prototypical model of the first class is the one- 
dimensional Ising model with ferromagnetic interactions decaying with distance $r$ as $r^{-2}$. It exhibits a 
Kosterlitz-Thouless (KT) vortex unbinding transition and it is dubbed IDSI for inverse distance squared Ising model. 
A paradigmatic model of the other class is the Poland-Scheraga model of DNA denaturation whereby the two strands of 
the DNA molecule separate from each other at a melting, or denaturation, temperature. It exhibits a condensation 
transition similar to the Bose Einstein condensation (BEC) transition of free bosons. In order to establish a link 
between these two classes of systems a modified version of the IDSI model has recently been introduced whereby the 
long-range interactions between the spins are restricted to exist only within domains of spins parallel to each 
other \cite{BM1, BM2, BMSM}. This model is dubbed TIDSI for truncated inverse distance squared Ising model. The 
model is exacly soluble and it exhibits a mixed order transition of the second class with algebraically diverging 
correlation length. The transition separates a totally ordered, non-fluctuating ferromagnetic phase from a 
disordered phase. Specific properties of this model have been studied and characterized, including the partition 
function, the distributions of cluster sizes, and the distribution of the length of the longest cluster.


We show below that part of the critical line of the TIDSI exhibits FDPO, 
characterized by extensive fluctuations and a cusp in the scaled 
correlation function as in Eq. (\ref {Eq_Cusp}). A key parameter in the 
model is the ratio of the strength of the inverse squared interaction to 
the temperature, denoted by $c$. Recent work on the distribution of the 
length $l_{\rm max}$ of the largest domain in various phases of this model 
has revealed~\cite{BMSM} that along the critical line and for $1<c<2$, 
while a typical domain is subextensive, the maximal domain is extensive. Moreover,
the exact form of the distribution~\cite{BMSM} of $l_{\rm max}$ on the critical line
indicates the existence of large fluctuations for $1<c<2$. This naturally raises the possibility
that on the critical line and for $1<c<2$, the TIDSI model may exhibit FDPO.
One good test would be to see if the spin-spin correlation function also
exhibits the signature of FDPO in this regime. In this paper,
we calculate exactly the spin-spin correlation functions $G(r|L)$ for the TIDSI model 
and show that along the critical line, there is a change from normal 
critical behaviour $G(r) \approx A/r^{c-2} \text{~~as~~} \, r 
\rightarrow \infty$ for $c>2$, to a size-dependent scaling function with a cusp 
singularity in the region $1<c<2$. Our result thus demonstrates very clearly
that indeed the region $1<c<2$ on the critical line exhibits FDPO.
	

The occurrence of FDPO in the TIDSI model brings out several interesting points. 
First, the TIDSI is an equilibrium system in contrast to the 
nonequilibrium systems studied in \cite{DB,DBM,KBB,CB}. It appears that 
the long-range interaction in the TIDSI model is the key element which induces 
FDPO, suggesting that FDPO may well occur in other settings where 
interactions are sufficiently long-ranged.  Secondly, we find that the 
cusp exponent $\alpha$ varies {\it continuously} along the critical line, 
as a function of a parameter. Such a variation of the cusp exponent has 
not been observed in earlier studies of FDPO, within a single model. 
Thirdly, the onset of FDPO coincides with the point at which the maximal 
domain becomes extensive. This interesting correlation between FDPO and 
extreme value statistics has been observed before in the context of a 
coarse-grained depth (CD) model \cite{DBM,CB} which mimics particles 
sliding down fluctuating surfaces in the adiabatic limit. Lastly, our 
study brings into focus the general question of the relation between FDPO 
and MOTs, as there are several other examples of MOTs which are associated 
with FDPO.

The remainder of the paper is organized as follows. In Section II, we 
define the TIDSI model, and show that when domain sizes are large, for 
instance near the critical line, one may represent configurations in terms 
of domains. In Section III we compute the asymptotic behavior of the partition function and 
derive the phase diagram. The Section also contains the computation of the
marginal domain size distribution in different regions of the phase boundary. Section IV 
discusses FDPO in the TIDSI model in terms of the 
two-point spin-spin correlation function. In the concluding Section IV, we discuss 
the issues set out in the previous paragraph, along with some open 
questions. Some details and an alternative derivation of the correlation function
are presented in Appendix A.

\section{The Model and the Domain Representation}

The TIDSI model is an Ising model defined on a one-dimensional lattice 
where on each site $i$ there is a spin variable $\sigma_i = \pm 1$. The 
interaction between spins is composed of a nearest neighbor term 
$-J_{NN}\sigma_i \sigma_{i+1}$ together with a long-range interaction term 
$-J(i-j)\sigma_i \sigma_j I(i \sim j)$, where $I(i \sim j)=1$ as long as 
sites $i$ and $j$ are in the same domain of either all up or all down 
spins and $I(i \sim j)=0$ otherwise. The long-range coupling is taken to 
be of the form
\begin{equation} 
J(r) \approx \frac{C}{r^2} \quad\, \text{for} \, \, r\gg1
\label{Jr.1}      
\end{equation}
The indicator function $I(i \sim j)$ may be expressed in terms of the spin 
variables
\begin{equation}
I(i \sim j)=\prod^{j-1}_{k=i} \delta_{\sigma_k \sigma_{k+1}}=
\prod^{j-1}_{k=i} \frac{1+\sigma_k \sigma_{k+1}}{2}
\end{equation}
The TIDSI Hamiltonian may thus be written as
\begin{equation}
\mathcal{H} =-J_{NN} \sum_{i=1}^{N} \sigma_i \sigma_{i+1} - 
\sum_{i<j} J(i-j)   \sigma_i \sigma_j \prod^{j}_{k=i} 
\frac{1+\sigma_k \sigma_{k+1}}{2}
\end{equation}

\begin{figure}
\includegraphics[width=1.0\textwidth]{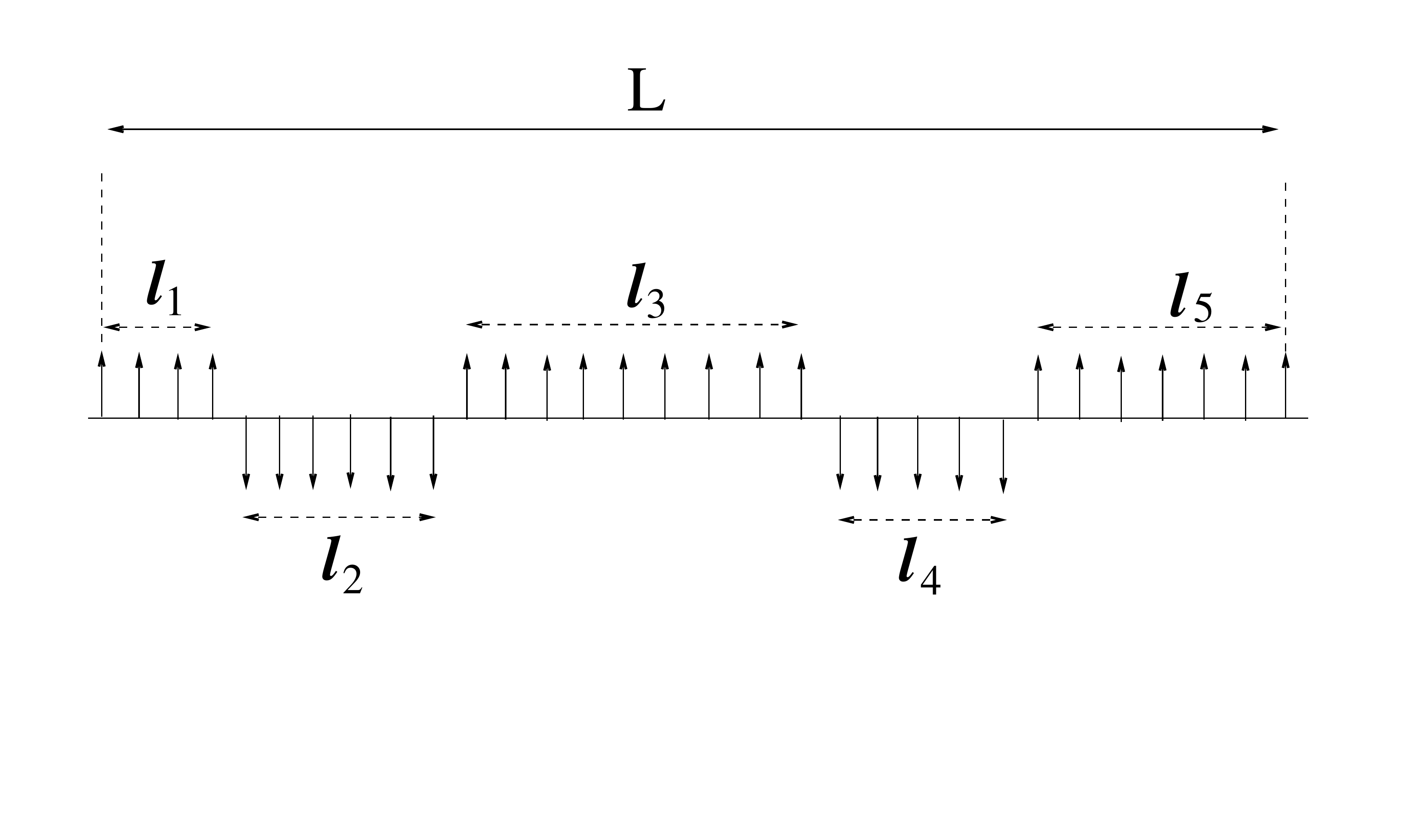}
\caption{A typical configuration of the domains in the TIDSI 
model, where the number of domains $N=5$ with domain lengths
$l_1$, $l_2$, $l_3$, $l_4$ and $l_5$. They satisfy the sum rule
$\sum_{i=1}^N l_i= L$ with $L$ denoting the system size.}
\label{fig_domain}
\end{figure}

It is convenient to express the Hamiltonian in terms of the domain length 
representation, where a domain is defined as a stretch of successive 
parallel spins (see Fig.~\ref{fig_domain}). This representation has been 
described in (\cite{BM2,BMSM}) but 
a brief account is included here for completeness. The long-range 
interaction in the second term operates only between pairs of spins which 
belong to the same domain, while the nearest neighbor interaction in the 
first term results in an energy cost for each domain wall. A typical 
configuration $\mathcal{C}$ is thus described by a set of domains with 
lengths $\{l_1,\, l_2,\, \cdots,\, l_N \}$ where the number of domains $N$ can 
vary from one configuration to another. The total system size $L$ and 
Hamiltonian can be expressed as

\begin{equation}        
\sum^N_{n=1} l_n = L
\end{equation}
\begin{equation}       
\mathcal{H} = \sum^N_{n=1} \mathcal{H}_n - J_{NN}
 \end{equation}
where
\begin{equation}       
\mathcal{H}_n = -J_{NN}(l_n-2) -  \sum^{l_n}_{r=1}(l_n-r)J(r)
 \end{equation}
Using the form of $J(r)$ in Eq. (\ref{Jr.1}), one can estimate
the sum via replacing it by an integral as
\begin{eqnarray}
\sum J(r) &\approx & a_0 - \frac{C}{l_n} \nonumber \\
\sum rJ(r)& \approx & b_0 + C \, \ln (l_n)\, ,
\label{sum.1}
\end{eqnarray}
where we have assumed $l_n$ is large and kept the two leading order
terms for large $l_n$. This 
is justified
since we are interested in phenomena close to the critical 
line where domains are typically large.  
Dropping an overall unimportant constant,
one obtains an the effective Hamiltonian 
\begin{equation} 
\mathcal{H} = C \sum_n \ln(l_n) + \Delta N
\end{equation}
where the constant $C$ is the amplitude of the long-range interaction and 
$\Delta = 2J_{NN}+C + b_0$ acts as a chemical potential for 
the number of domains. It is useful to define the 
parameter $c=\beta C$, as it enters in an important way in the 
subsequent development. 

As mentioned before, a configuration $\mathcal{C}$ of the system
is now specified by the domain sizes, as well as the number of domains 
$N$: $\mathcal{C} \equiv \{l_1,\, l_2,\, \cdots, \,l_N,\, N \}$.
The probability of such a configuration $\mathcal{C}$ is
given by its Boltzmann weight
\begin{equation}        
P(l_1,\, l_2,\, \cdots,\, l_N,\, N|L) = 
\frac{y^N}{Z_y(L)}\, \prod^N_{n=1}\, 
\frac{1}{l_n^c} \, \delta_{\sum^N_{n=1}l_n, \, L }  
\label{jointd.1}
\end{equation}
where $y=e^{-\beta \Delta}$ and $\delta_{i,j}$ is the Kronecker
delta function that enforces the sum rule. The normalization constant
$Z_y(L)$ is indeed
the partition function given by 
\begin{equation}        
Z_y(L) =  \sum^\infty_{N=1} y^N \sum^\infty_{l_1=1} 
\ldots \sum_{l_N=1}^{\infty} \prod^N_{n=1} 
\frac{1}{l_n^c} \, \, \, \delta_{\sum^N_{n=1} l_n, \,L} \, .
\label{pf.1}
\end{equation}

As we will see below, the way in which $Z_y(L)$ scales with system size 
$L$ changes depending on the value of the two parameters $c$ and $y$. We 
will thus consider
$c$ and $y$ as independent parameters and discuss the behaviour of the system in 
different regimes in the $(c-y)$ plane. Even though the joint distribution
in Eq. (\ref{jointd.1}) is well defined for any $c>0$, it turns out
that for $0<c\le 1$, there is no phase transition as a function of $y$ and
the system is always in a paramagnetic phase. In contrast, for $c>1$, 
there is a phase transition in the $(c-y)$ plane across the critical
line $y_c=1/\zeta(c)$ where $\zeta(c)= \sum_{l=1}^{\infty} l^{-c}$ is
the Riemann zeta function. For $c>1$, the system is in a 
paramagnetic phase for $y>y_c$, while it is ferromagnetic for $y<y_c$ (see the phase diagram
in Fig. (\ref{fig_phasediagram})).
Hence, in the rest of the paper, we will restrict ourselves to the
case $c>1$.

\section{Partition function, phase diagram and domain size distribution}

The partition function and the phase diagram of this model has been 
analysed before in the $(c-T)$ plane in Refs~\cite{BM1,BM2}. In this 
section, we re-derive some of these results in the $(c-y)$ plane (the 
details are slightly different from those in the $(c-T)$ plane). Some of 
these results will be useful later for computing the marginal domain size 
distribution, as well as the spin-spin correlation function.

\subsection{Partition function and phase diagram}

To analyse the behaviour of  the partition function $Z_y(L)$ in Eq. 
(\ref{pf.1}) in different regimes in the $(c-y)$ plane, let us 
define its generating function
\begin{equation}        
\tilde{Z}_y(s) =  \sum^{\infty}_{L=1} e^{-sL} Z_y(L) \, .
\label{pfgf.1}
\end{equation}
The generating function corresponding to  Eq. (\ref{pf.1}) is
\begin{equation}        
\tilde{Z}_y(s) = \sum^{\infty}_{L=1} e^{-sL} Z_y(L)=   \frac{y \phi(s)}{1- 
y \phi(s)}
\label{pfgf.2}
\end{equation}
with
\begin{equation}        
\phi(s) =  \sum^{\infty}_{l=1} \frac{e^{-sl}}{l^c}= {\rm 
Li}_c\left(e^{-s}\right)\, ,
\label{phis.1}
\end{equation}
where ${\rm Li}_c(z)= \sum_{l=1}^{\infty} \frac{z^l}{l^c}$ is the 
polylogarithmic function. Thus, to extract the large $L$ asymptotic 
behavior of $Z_y(L)$, we need to analyse the singularities of the
right hand side (rhs) of Eq. (\ref{pfgf.2}) as a function of $s$.

Clearly $\phi(s)$ in Eq. (\ref{phis.1}) decreases monotonically as
$s$ increases from $0$ to $\infty$, starting from $\phi(0)= \zeta(c)$.
Near $s=0$, using the known asymptotic behavior of polylogarithms, one can
show that $\phi(s)$ has the 
asymptotic expansion
\begin{equation}
\phi(s)= \sum_{k=0}^{n-1} \frac{(-s)^k}{k!} \zeta(c-k) + \Gamma(1-c)\, 
s^{c-1}+\dots
\label{phis_asymp0.1}
\end{equation}
where $n={\rm int}[c]$ (for integer $c$, the first non-analytic term
gets additional multiplicative logarithmic corrections). In contrast, as $s\to \infty$, the leading 
behavior of $\phi(s)$ comes from the $l=1$ term in Eq. (\ref{phis.1}), implying $\phi(s) \approx e^{-s}$ for large $s$. Thus,
as a function of $s$, the rhs of Eq. (\ref{pfgf.2}) has a pole
at some $s=s^*>0$, provided $1/y<\phi(0)=\zeta(c)$. We will see later that this corresponds to
the paramagnetic phase. As $y\to y_c=1/\zeta(c)$ from above, the pole
$s^*\to 0$ and we need to analyse the rhs for small $s$ and we will
see below that $y<y_c$ will correspond to the ferromagnetic phase.
Below we analyse the large $L$ behavior of $Z_y(L)$ in the three
regimes separately: (i) paramagnetic phase ($y>y_c$) (ii) critical point 
($y=y_c$) and (iii) ferromagnetic phase ($y<y_c$).

Since the large $L$ behavior of $Z_y(L)$ corresponds to the small $s$ behavior of the generating function ${\tilde Z}_y(s)$, 
we first approximate, for small $s$, the sum in Eq. (\ref{pfgf.1}) 
by an integral, i.e., the generating function coincides with the Laplace transform
\begin{equation}
Z_y(s) \approx \int_0^{\infty} Z_y(L)\, e^{-sL}\, dL = \frac{y \phi(s)}{1-y \phi(s)}\, .
\label{pflt.1}
\end{equation}
Inverting the Laplace transform, we can express $Z_y(L)$ as a Bromwich integral in the complex $s$ plane
\begin{equation}
Z_y(L)= \int_{\Gamma_0} \frac{ds}{2\pi i}\, e^{sL}\, \frac{y \phi(s)}{1-y \phi(s)}
\label{brom.1}
\end{equation}
where $\Gamma_0$ is a vertical contour whose real part is to the right of all singularities of the 
integrand in the complex $s$ plane. Below
we analyse the large $L$ behavior of this Bromwich integral in the three regimes. 

\vskip 0.2cm

\noindent {\bf {(i) Paramagnetic phase ($y>y_c$):}} In this case, the integrand in Eq. (\ref{brom.1}) has a pole
at $s=s^*>0$, where $y\phi(s^*)=1$. As argued above, this happens provided $y>y_c=1/\phi(0)=1/\zeta(c)$. In this case,
the leading large $L$ behavior of the Bromwich integral comes from this pole $s^*$ in the $s$ plane. Evaluating the residue, we obtain
\begin{equation}
Z_y(L) \approx B_0\, e^{s^* L} \, ; \quad {\rm where} \quad B_0=\frac{1}{-y\phi'(s^*)}
\label{pf_para.1}
\end{equation}
Thus the free energy $-\ln Z_y(L) \sim s^* L$ is extensive in $L$, clearly indicating that we are in a paramagnetic phase.
As $y\to y_c$ from above, for fixed $c>1$, the pole $s^*\to 0$, and we need to analyse the 
nonanalytic behavior of the
integrand near its branch cut at $s=0$.

\vskip 0.2cm

\noindent {\bf {(ii) Critical line ($y=y_c$):}} We set $y=y_c$ on the rhs of Eq. (\ref{pflt.1}) and replace $\phi(s)$ by its small $s$  
behavior in Eq. (\ref{phis_asymp0.1}). For the leading $s$ behavior, in the numerator $y_c\phi(s)$ on the rhs in Eq. (\ref{pflt.1}), 
we can just keep the leading term
$\phi(s)= \zeta(c)$ (the higher order corrections lead to only subleading behavior). Hence $y\phi(s) \approx 1$. In contrast, 
using $y_c=1/\zeta(c)$, the leading term for small $s$
in the denominator depends crucially on whether $c>2$ or $1<c<2$.

\vskip 0.2cm

\begin{itemize}

\item $c>2$: For $c>2$, the leading order term in the denominator of the rhs of Eq. (\ref{pflt.1}) is the analytic term,  
$1-y_c\phi(s) \approx y_c \zeta(c-1) s$. Hence,
$Z_y(s) \approx 1/[y_c \zeta(c-1) s]$, whose Laplace inversion gives trivially
\begin{equation}
Z_y(L) \approx \frac{1}{y_c \zeta(c-1)} = \frac{\zeta(c)}{\zeta(c-1)}=B_1 \, .
\label{pf_crit1.1}
\end{equation}
Thus, the partition function approaches a constant $B_1$ as $L\to \infty$.
\item $1<c<2$: In this case, the leading order term in the denominator of the rhs of Eq. (\ref{pflt.1}) 
for small $s$ is the non-analytic term in the small $s$ expansion of $\phi(s)$
in Eq. (\ref{phis_asymp0.1}), i.e., $1-y_c \phi(s) \approx -y_c \Gamma(1-c)\, s^{c-1}$. Hence,
$Z_y(s) \approx - 1/[y_c \Gamma(1-c) s^{c-1}]$ as $s\to 0$. Inverting the Laplace transform in a straightforward way
and simplifying, we get
\begin{equation}
Z_y(L) \approx B_2\, L^{c-2}\, ; \quad {\rm where} \quad B_2= \frac{1}{\pi}\zeta(c) (c-1) \sin(\pi(c-1))\, .
\label{pf_crit2.1}
\end{equation}
Thus, for $1<c<2$, the partition function decays algebraically for large $L$ as $L^{c-2}$.

\end{itemize}

\vskip 0.2cm

\noindent {\bf {(iii) Ferromagnetic phase ($y<y_c$):}} Finally, we turn to the ferromagnetic phase $y<y_c=1/\zeta(c)$. 
In this case, substituting
the small $s$ behavior of $\phi(s)$ from Eq. (\ref{phis_asymp0.1}) in the rhs of Eq. (\ref{pflt.1}), we find the following leading
small $s$ behavior for the Laplace transform
\begin{equation}
\int_0^{\infty} Z_y(L)\, e^{-sL}\, dL\approx  
\frac{y\zeta(c)}{1-y \zeta(c)} + ``{\rm analytic\,\, terms}'' + \frac{y\Gamma(1-c)}{(1-y\zeta(c))^2}\, s^{c-1} + \cdots
\label{pf_ferro_lt.1}
\end{equation}
Note that the leading constant term is positive if and only if $y<y_c=1/\zeta(c)$, clearly indicating that this
expansion makes sense only in the ferromagnetic phase.
The leading nonanalytic term for small $s$ in the Laplace transform in Eq.({\ref{pf_ferro_lt.1}) fixes the leading large $L$ 
behavior of $Z_y(L)$ uniquely via
a Tauberian theorem and we get 
\begin{equation}
Z_y(L) \approx B_3\, L^{-c}\, ; \quad {\rm where} \quad B_3= \frac{y}{(1-y\zeta(c))^2}\, .
\label{pf_ferro.1}
\end{equation}
Thus, in the ferromagnetic phase, for any $c>1$, the partition function decays as $L^{-c}$ for large $L$.

Let us then just summarize the behavior of the partition function $Z_y(L)$ for large $L$ in the $(c-y)$ plane
in Fig. (\ref{fig_phasediagram}):
\begin{eqnarray}
Z_y(L) \approx \begin{cases}

& B_0\, e^{s^*\, L}\quad {\rm for} \quad y>y_c=1/\zeta(c) \quad (\rm PARA)   \\
& B_1 \quad\quad \quad\, {\rm for} \quad y=y_c\,\, {\rm and}\,\, c>2 \quad (\rm CRITICAL\,\, LINE) \\
& B_2\, L^{c-2} \quad {\rm for} \quad y=y_c\,\, {\rm and}\,\, 1<c<2 \quad (\rm CRITICAL\,\, LINE)\\
& B_3\, L^{-c} \quad\,\, {\rm for} \quad y<y_c=1/\zeta(c) \quad (\rm FERRO)   \\
\end{cases}
\label{pf_summary}
\end{eqnarray}
where $s^*$ is the solution of $y\phi(s^*)=1$ for $y>y_c$ and the four constants $B_0$, $B_1$, $B_2$ and $B_3$ are given respectively in Eqs. (\ref{pf_para.1}), (\ref{pf_crit1.1}),
(\ref{pf_crit2.1}) and (\ref{pf_ferro.1}). 
This also leads to the phase diagram in Fig. (\ref{fig_phasediagram}). We emphasize that on the 
critical line $y=y_c$, the partition function behaves rather differently as a function of $L$ for $c>2$ and
$1<c<2$ (shown respectively by the dashed line and the soild (red) line in the phase diagram
in Fig. (\ref{fig_phasediagram}).

\begin{figure}
\includegraphics[width=0.6\textwidth]{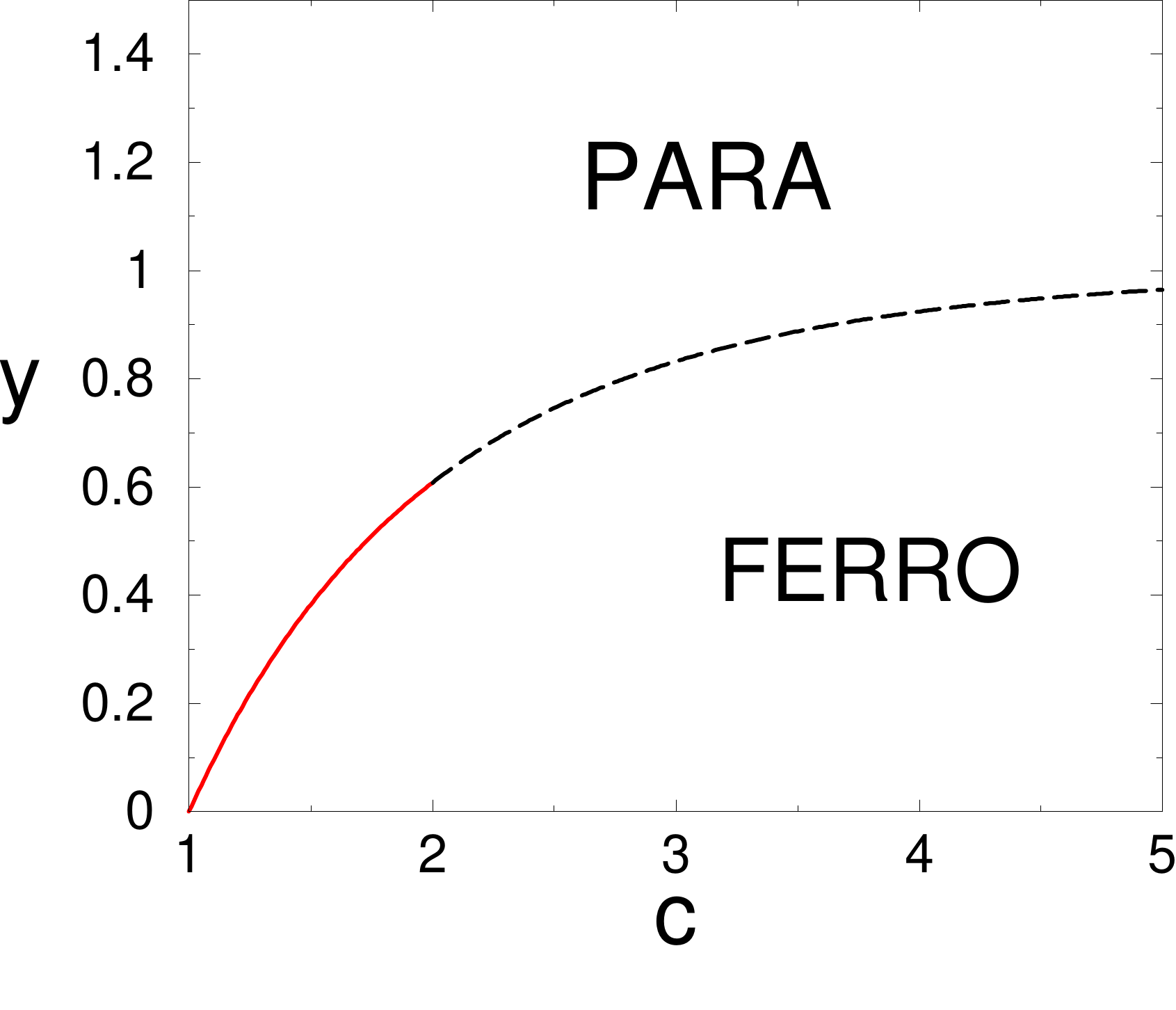}
\caption{ Phase diagram in the $(c-y)$ plane, where $c> 1$.
The critical line $y_c=1/\zeta(c)$ separates the paramagnetic
phase ($y>y_c$) from the ferromagnetic phase ($y<y_c$).
The system exhibits FDPO on the part
$1<c\le 2$ of the critical line (shown by the solid (red) line).}
\label{fig_phasediagram}
\end{figure}

\subsection{Distribution of the domain sizes}

In this subsection, we compute the marginal domain size distribution $P_y(l|L)$, i.e.,
the probability that a randomly picked domain has size $l$, given the total system size $L$ and for fixed $c$ and $y$.
This is done as follows. We start from the joint distribution of domain lengths and the number of domains in
Eq. (\ref{jointd.1}), keep one of the domain lengths fixed at $l$ (say $l_1=l$), and sum over all other $l_i$'s 
as well as $N$. This gives
\begin{equation}
P(l|L)= \frac{y}{l^c}\, \frac{1}{Z_y(L)}\, \sum_{N=1}^{\infty}\, \sum_{l_2\ge 1,\cdots, l_N\ge 1} \frac{y^{N-1}}{l_2^c\,l_3^c
  \cdots l_N^c}\, \delta_{l_2+l_3+\cdots +l_N, L-l}\, .
\label{Pl.1}
\end{equation}
Note that, using the partition function $Z_y(L)$ in Eq. (\ref{pf.1}), the marginal distribution
satisfies, by construction, the normalization condition $\sum_{l\ge 1} P(l|L)=1$. Indeed, $P(l|L)$ can also be
interpreted as follows: given that a domain wall occurs, $P(l|L)$ is the probability that the
next domain wall to the right occurs at a distance $l$. 

To proceed, we consider the sum over $N=1,2\cdots$ in Eq. (\ref{Pl.1}), 
and separate the $N=1$ term (only one domain in the whole system) and $N\ge 2$ terms. The sum over 
$N\ge 2$ can be 
reexpressed in terms of the partition function $Z_y(L)$ in Eq. (\ref{pf.1}). This gives
\begin{equation}
P(l|L)= \frac{y}{L^c\,Z_y(L)}\, \delta_{l,\,L} + \frac{y}{l^c}\, \frac{Z_y(L-l)}{Z_y(L)}\,
\label{Pl.2}
\end{equation}
where the first term corresponds to $N=1$. Note that Eq. (\ref{Pl.2})
is exact for all $l\ge 1$ and $L\ge 1$. The next step is to analyse this marginal distribution $P(l|L)$ for large $L$ in
different regions of the phase diagram in the $(c-y)$ plane in Fig. (\ref{fig_phasediagram}). For this, we will use
the asymptotic properties of the partition function $Z_y(L)$ derived in the previous subsection that are 
summarized in Eq. (\ref{pf_summary}).

\vskip 0.2cm

\noindent {\bf {(i) Paramagnetic phase ($y>y_c=1/\zeta(c)$):}} In this regime, $Z_y(L) \sim e^{s^*\, L}$ for large $L$ from Eq. 
(\ref{pf_summary}), where $s^*>0$ is the root of $y\,\phi(s^*)=1$ with $y>y_c$ (recall that $\phi(s)$ is given in
Eq. (\ref{phis.1})). The first term in Eq. (\ref{Pl.2}), i.e., the delta peak behaves as
$\approx \frac{y}{L^c}\, e^{-s^*\, L}\, \delta_{l,L}$. Hence, its amplitude vanishes exponentially as $L\to \infty$
and hence this first term can be dropped in the thermodynamic limit. In the second term in Eq. (\ref{Pl.2}),  
assuming $L-l\gg 1$ in the numerator, we get
\begin{equation}
P(l|L) \approx  \frac{y}{l^c}\, \frac{Z_y(L-l)}{Z_y(L)} \approx \frac{y}{l^c}\, e^{-s^*\, l}\, .
\label{Pl_para.1}
\end{equation}
Hence, as $L\to \infty$, we obtain a domain size distribution
independent of $L$
\begin{equation}
P(l|L) \approx \frac{y}{l^c}\, e^{-s^*\, l}
\label{Pl_para.2}
\end{equation}
that has an exponential tail for large $l$. Consequently, the average domain length is a constant of $O(1)$ in
the thermodynamic limit $L\to \infty$ and is given by 
\begin{equation}
\langle l\rangle = \sum_{l=1}^L l\, P(l|L) \approx  \sum_{l=1}^{\infty} \frac{y}{l^{c-1}}\, e^{-s^*\, l} \sim O(1)\, .   
\label{avgl_para.1}
\end{equation}

\vskip 0.2cm

\noindent {\bf {(ii) Critical line ($y=y_c$):}} We have seen before that the partition function on the critical
line $y=y_c$ behaves differently for $c>2$ and $1<c<2$ (see Eq. (\ref{pf_summary}). Consequently, the domain size 
distribution $P(l|L)$ on the critical line also has different behaviors respectively for $c>2$ and $1<c<2$. Below, 
we consider these two cases separately.

\vskip 0.2cm

\begin{itemize} 

\item $c>2$: In this case, for large $L$, $Z_y(L) \approx B_1$ from Eq. (\ref{pf_summary}), 
where $B_1= \zeta(c)/\zeta(c-1)$ is a constant. Substituting this
in Eq. (\ref{Pl.2}), the first term behaves as $\approx \frac{y_c}{L^c B_1}\, \delta_{l,\, L}$. Thus the
amplitude of the delta peak again vanishes as $L\to \infty$, albeit algebraically. Dropping this term, 
assuming $L-l\gg 1$ in the numerator of the second term in Eq. (\ref{Pl.2}) we get a power law distribution, 
independent of $L$ for large $L$ 
\begin{equation}
P(l|L) \approx \frac{y_c}{l^c}\, \frac{Z_y(L-l)}{Z_y(L)}\approx \frac{y_c}{l^c}\, .
\label{Pl_ucrit.2}
\end{equation}
Thus the distribution $P(l|L)$ has the same power law tail as the L\'evy stable distribution ${\cal L}_\mu(l)$ with L\'evy index $\mu=c-1>1$.
Consequently, the average domain length is finite, i.e., of $O(1)$ as $L\to \infty$
\begin{equation}
\langle l\rangle = \sum_{l=1}^L l\, P(l|L) \approx y_c \sum_{l=1}^{\infty} \frac{1}{l^{c-1}}= y_c\, \zeta(c-1)\sim  O(1)\, .
\label{avgl_ucrit.1}
\end{equation}

\item $1<c<2$: In this case, $Z_y(L) \approx B_2\, L^{c-2}$ from Eq. (\ref{pf_summary}), where $B_2$ is a constant given in
Eq. (\ref{pf_crit2.1}). Substituting this in Eq. (\ref{Pl.2}), the first term behaves as
$\approx \frac{y_c}{B_2\, L^{2(c-1)}}\, \delta_{l, \, L}$. Once again, the amplitude of the delta peak
decays algebraically as $L^{-2(c-1})$ for large $L$ and vanishes in the thermodynamic limit. 
Furthermore, 
assuming that we can use this asymptotic form of $Z_y(L)$ also
in the numerator $Z_y(L-l)$ in the second term of Eq. (\ref{Pl.2}), we get
\begin{equation}
P(l|L) \approx \frac{y_c}{l^c}\, \frac{Z_y(L-l)}{Z_y(L)}\approx \frac{y_c}{l^c} \left(1- \frac{l}{L}\right)^{c-2}\, .
\label{Pl_lcrit.2}
\end{equation}
Note that there is still a nontrivial $L$ dependence in $P(l|L)$ even for large $L$--the distribution still depends
on $L$ for $l\sim L$. As $L\to \infty$, the part of the distribution for $l<<L$ does become independent of $L$
\begin{equation}
P(l|L) \sim \frac{y_c}{l^c}\, ; \quad {\rm for}\quad l<< L \, .
\label{Pl_lcrit.3}
\end{equation}
However, when $l$ approaches its upper cut-off $L$, the distribution $P(l|L)$ diverges as $(1-l/L)^{c-2}$, though it still 
remains integrable. As a result, while the distribution itself converges to a power law form as in Eq. (\ref{Pl_lcrit.3})
for $l\ll L$, all its moments (including the average) diverges algebraically with $L$ as $L\to \infty$. 
This is because, the moments are dominated by contributions coming from the upper cut-off region $l\sim L$. For example, 
the average domain size, using Eq. (\ref{Pl_lcrit.3}) behaves as
\begin{equation}
\langle l\rangle = \sum_{l=1}^L l\, P(l|L) \approx \frac{y_c}{2-c} L^{2-c}\, .
\label{avgl_lcrit.1}
\end{equation}
Similarly, all higher moments also diverge algebraically. Thus, for large $L$, the distribution
$P(l|L)$ decays with the same power as the L\'evy stable distribution ${\cal L}_\mu(l)$ with
L\'evy index $0<\mu=c-1<1$, for which all positive integer moments diverge, even though the distribution 
itself is normalizable. We will see later that this strong fat tail of the domain size 
distribution for the $1<c<2$ case,
with moments diverging with increasing $L$ (leading to extremely large fluctuations), also affects the
$L$ dependence of the spin-spin correlation in a manner consistent with the FDPO scenario.

\end{itemize}

\vskip 0.2cm

\noindent {\bf {(iii) Ferromagnetic phase ($y<y_c$):}} In this phase, from Eq. (\ref{pf_summary}), we have 
$Z_y(L) \approx B_3\, L^{-c}$ for large $L$, where $B_3= y/(1-y\zeta(c))^2$ from Eq. (\ref{pf_ferro.1}).
We recall that in this phase $y<y_c= 1/\zeta(c)$.
Substituting this behavior in the first term of
Eq. (\ref{Pl.2}), we find that, in contrast to the para phase or the critical line, the amplitude of
the delta peak approaches a constant $y/B_3=(1-y\zeta(c))^2= (1-y/y_c)^2$, as $L\to \infty$--this is the typical signature of the ferromagnetic phase where
with a nonzero probability the system has one single domain of size $L$. This is akin to the condensation
phenomenon where a single term $l=L$ carries a finite fraction of the probability weight, leading to an ordered state. 
Hence we get
\begin{equation}
P(l|L) \approx \left(1-\frac{y}{y_c}\right)^2\, \delta_{l,L} + \frac{y}{l^c}\, \frac{Z_y(L-l)}{Z_y(L)}\, .
\label{Pl_ferro.1}
\end{equation}
When $y\to y_c$ from below, the amplitude of the delta peak vanishes. Since $P(l|L)$ is normalized to 
unity, the
non-delta peak part carries a total weight of $1- (1-y/y_c)^2$. Now, for $l\ll L$, this second term can be approximated
by substituting $Z_y(L) \approx B_3\, L^{-c}$ and taking $L\to \infty$ limit leads to a power law tail
\begin{equation}
\frac{y}{l^c}\, \frac{Z_y(L-l)}{Z_y(L)}\approx \frac{y_c}{l^c}\, ; \quad {\rm for}\quad l\ll L
\label{Pl_ferro.2}
\end{equation}
In the regime where $L-l \sim O(1)$, it is a bit complicated to estimate the precise form of $P(l,L)$. Thus summarizing, 
in the ferro phase, the distribution has a (i) power law part, $P(l|L) \sim y\, l^{-c}$ for $l<< L$, (ii) has a genuine
delta peak at its upper cut-off $l=L$, i.e., $P(l=L|L)= (1-y/y_c)^2$ and (iii) has a nontrivial form in
the intermediate regime $1<< l < L$, i.e., when $L-l \sim O(1)$. Note that when we sum over $l$, the third regime (iii)
contributes a finite amount to the normalization. These three regimes in the ferro phase are very similar to the 
distribution of the mass at a fixed site in the
well studied mass transport models such as the zero range process, 
in its condensed phase~\cite{MEZ05,EMZ06}. Finally, the average domain length is given by
\begin{equation}
\langle l\rangle = \sum_{l=1}^L l\,P(l|L)\approx \left(1-\frac{y}{y_c}\right)^2\, L  + O(L^{2-c})\, ,
\label{avgl_ferro.1}
\end{equation}
where the leading $\sim O(L)$ term comes from the delta peak at $l=L$, while the rest of the distribution 
contributes
to the subleading term $O(L^{2-c})$ (note that for any $c>1$, $L^{2-c}\ll L$ for large $L$). 

Summarizing, the average domain length scales with system size $L$ for large $L$ in the following manner in the four 
different regimes in the $(c-y)$ plane (see Fig. (\ref{fig_phasediagram}))
\begin{eqnarray}
\langle l\rangle  \sim \begin{cases}

& O(1) \quad {\rm for} \quad y>y_c=1/\zeta(c) \quad (\rm PARA)   \\
& O(1) \quad {\rm for} \quad y=y_c\,\, {\rm and}\,\, c>2 \quad (\rm CRITICAL\,\, LINE) \\
& L^{2-c} \quad {\rm for} \quad y=y_c\,\, {\rm and}\,\, 1<c<2 \quad (\rm CRITICAL\,\, LINE)\\
& L \quad\quad\,\,\, {\rm for} \quad y<y_c=1/\zeta(c) \quad (\rm FERRO)   \\
\end{cases}
\label{avgl_summary}
\end{eqnarray}
Consequently, the typical number of domains $N\sim L/{\langle l\rangle}$ scales as: $N\sim L$ (Para), $N\sim L$
(Critical line where $c>2$), $N\sim L^{c-1}$ (Critical line where $1<c<2$) and $N\sim O(1)$ (Ferro). Thus, both
in the para phase, as well as on the critical line where $c>2$, the number of domains is extensive. 
On the critical line where $1<c<2$, the number of domains still grows with $L$, but only
subextensively since $L^{c-1}<< L$ for large $L$. Finally in the ferro phase, condensation takes place
and the system essentially consists of a single large domain with size proportional to $L$.

We conclude this subsection with one final remark. In the discussion above, we have computed the marginal
size distribution of a single domain $P(l|L)$, i.e., the one point domain size distribution function. 
One can also compute, in a similar fashion, the marginal $m$-point domain size distribution
$P(l_1,\,l_2,\,\cdots,\, l_m)$ by keeping the sizes of $m\ge 1$ domains fixed at 
$\{l_1,\,l_2,\,\cdots,\, l_m\}$ and summing over the rest.
It is easy to see that both in the paramagnetic side ($y>y_c$) as well as on the critical line ($y=y_c$ and 
for any $c>1$), the
$m$-point size distribution factorises into a product of one-point distribution in the limit of large $L$
\begin{equation}
P(l_1,\,l_2,\,\cdots,\,l_m|L) \approx P(l_1|L)\,P(l_2|L)\cdots P(l_m|L)\, . 
\label{IIA.1}
\end{equation}
In other words, for $y\ge y_c$ (para phase and the critical line) the global constraint imposed by the 
delta function in the joint distribution in Eq. (\ref{jointd.1}) does
not induce any correlation between domains in the large $L$ limit, and the independent interval approximation (IIA)
becomes exact. In contrast, in the ferro phase $y<y_c$, this factorisation no longer holds as the system is 
essentially dominated
by a single large domain and the global constraint induces significant correlations between domains.

\section{FDPO in the TIDSI model via the spin-spin correlation function}

There are two principal hallmarks of the FDPO state, namely (a) correlations of the
order parameter which persist at a distance that scales with the system size $L$ and do not damp down
in the thermodynamic limit $L\to \infty$ and 
(b) a cusp singularity in the correlation function of the order parameter at small values of the scaled 
separation $r$ (when $r$ is small compared to $L$ but large compared to any microscopic scale, i.e. 
for $1\ll r\ll L$). In this section we show that both (a) and (b) are manifest in 
the TIDSI model, along the critical line in the region $1 < c < 2$.

Various quantities show anomalously large fluctuations in FDPO. Thus for 
instance, for the system of particles sliding down fluctuating surfaces, 
each of the multiple order parameters that characterize the FDPO state 
asymptotes to a broad distribution as the system size $L \rightarrow 
\infty$ \cite{KBB}. Likewise, $l_{\rm max}$, the length of the largest 
connected domain of particles scales with the system size $L$, and the 
corresponding probability distribution of the scaled variable 
$y=l_{max}/L$ approaches an asymptotic form in the thermodynamic limit 
\cite{DBM,CB}.

In the TIDSI model, the statistics of $l_{\rm max}$ has been studied in 
detail, and the corresponding probability density (PDF) has been derived 
in~\cite{BMSM}. The MOT involves a transition from a disordered state with 
multiple domains (where the centered and scaled distribution of $l_{\rm max}$ 
follows a Gumbel distribution), to an 
ordered state consisting of essentially one single macroscopically large domain. The distribution 
of $l_{\rm max}$ along the critical curve is interesting. 
For $c>2$, the 
PDF of $l_{\rm max}$ is a Fr\`echet distribution with argument $l_{\rm max}/ 
L^{1/(c-1)}$. But for $1 < c < 2$, the PDF is a function of $y=l_{\rm max}/L$. 
The corresponding scaling function approaches a broad limiting function of the ratio 
$l_{\rm max}/L$ which was found analytically and shown to have a succession 
of ever weakening singularities at a denumerable set of points 
\cite{BMSM}. That the limiting form of the distribution is not a delta 
function indicates FDPO.

Thus, along the critical line and for $1<c<2$ (shown the solid (red) line in
Fig. (\ref{fig_phasediagram})), the FDPO must be manifest also in
the spin-spin correlation function. We now demonstrate that indeed this is the
case by computing the spin-spin correlation function $G(r|L)= \langle \sigma_i \sigma_{i+r}\rangle$
along the critical line $y=y_c$. Consider two spins $\sigma_i$ and $\sigma_{i+r}$ separated by $r$ sites.
Since $\sigma_i=\pm 1$ for any $i$, in any spin configuration the product $\sigma_i\sigma_{i+r}$ is also either $+1$ (if there
are even number of domain walls between $i$ and $i+r$) or $-1$ (if there are odd number of domain
walls between $i$ and $i+r$). Consequently, taking the average over all spin configurations one can write
a very general exact expression
\begin{equation}
G(r|L)= \sum_{n=0}^{\infty} (-1)^n\, p_n(r|L)\, ,
\label{gencorr.1}
\end{equation}
where $p_n(r|L)$ denotes the probability of having exactly $n$ domain walls between $i$ and $i+r$. Now, along
the critical line $y=y_c$ where the IIA holds (see Eq. (\ref{IIA.1})), i.e., when the domains
are statistically independent, then for $r\ll L$ and large $L$, one can show that the dominant contribution
to $G(r|L)$ in the sum in Eq. (\ref{gencorr.1}) comes from the $n=0$ term and the terms $n\ge 1$ provide only 
subleading corrections for large $L$ (see Appendix A). Hence, the
correlation function in this regime, for large $L$, can be well approximated by
\begin{equation}
G(r|L) \approx p_0(r|L)\, . 
\label{corr.2}
\end{equation}
Thus, we need to estimate $p_0(r|L)$, i.e., the probability that a random selected interval of size $r$ is
free of any domain wall. In other words, $p_0(r|L)$ is just the probability that both sites $i$ and $i+r$ belong to
the same domain.

Now $p_0(r|L)$ can be estimated in terms of the marginal domain size distribution $P(l|L)$ derived in the
previous section. To see this, we first compute the probability ${\cal P}_0(l|L)$ that
a randomly selected site $i$ belongs to a domain of size $l$. This is simply given in terms
of the domain size distribution $P(l|L)$ by the following relation
\begin{equation}
{\cal P}_0(l|L) = \frac{1}{\langle l\rangle}\, l\, P(l|L)\, ,
\label{single_site.1}
\end{equation}   
where $\langle l \rangle= \sum_{l=1}^L l P(l|L)$.
This is easily understood. The chosen site may be any one of the $l$ sites of a domain of size $l$ explaining
the factor $l$ multiplying $P(l|L)$, and the overall factor $1/\langle l\rangle$ ensures that ${\cal P}_0(l|L)$
is normalized to unity: $\sum_{l=1}^{L} {\cal P}_0(l|L)=1$. In the previous section we have estimated both
$P(l|L)$ as well as $\langle l\rangle $ (see Eq. (\ref{avgl_summary})). Hence, we have a precise estimate of
${\cal P}_0(l|L)$ for large $L$ in all regimes of the phase diagram in the $(c-y)$ plane. 
Given the probability ${\cal P}_0(l|L)$ that a randomly selected site belongs to a domain
of size $l$, the conditional probability that a site at a distance $r$ falls within the same domain of size $l$
is simply the ratio $\frac{(l-r)}{l}$. The latter is just the probability that a stick of size $r$ fits fully
within a domain of size $l$. Thus multiplying and summing over $l$ from $r$ to $L$, the probability 
$p_0(r|L)$ that sites $i$ and $i+r$ both belong to the same domain is given by
\begin{equation}
p_0(r|L) = \sum_{r}^L {\cal P}_0(l|L)\, \frac{l-r}{l} = \frac{1}{\langle l\rangle} \sum_{l=r}^L P(l|L) (l-r)\, ,
\label{p0r.1}
\end{equation}
where in establishing the last equality, we used Eq. (\ref{single_site.1}). Next we use the result from the
previous section that on the critical line $y=y_c$, $P(l|L) \approx y_c/l^c$ for all $c>1$ and $l\ll L$ (see
e.g. Eq. (\ref{Pl_ucrit.2}) for $c>2$ and Eq. (\ref{Pl_lcrit.3}) for $1<c<2$). This gives,  
\begin{equation}
G(r|L)\approx p_0(r|L) \approx \frac{y_c}{\langle l\rangle} \sum_{l=r}^L \frac{(l-r)}{l^c}\, .
\label{Gr.1}
\end{equation}

To estimate the sum in Eq. (\ref{Gr.1}), we consider $r\gg 1$ which enables us to replace the sum by an integral
\begin{equation}
G(r|L) \approx \frac{y_c}{\langle l\rangle} \int_r^L \frac{l-r}{l^c} \, dl\, ,
\label{Gr.2}
\end{equation}
that can be performed easily giving
\begin{equation}
G(r|L) \approx \frac{y_c}{\langle l\rangle}\, \left[ \frac{L^{2-c}}{2-c} - \frac{r^{2-c}}{(c-1)(2-c)} + 
\frac{r\,L^{1-c}}{c-1}\right]\, .
\label{Gr.3}
\end{equation}
Since $c>1$, we can drop the last term $\sim L^{c-1}$ for large $L$ for any $c>1$, leading to
\begin{equation}
G(r|L) \approx \frac{y_c}{\langle l\rangle}\left[ \frac{L^{2-c}}{2-c} - \frac{r^{2-c}}{(c-1)(2-c)}\right]\, .
\label{Gr.4}
\end{equation} 
We now show that $G(r|L)$ for large $L$ in Eq. (\ref{Gr.4}) behaves very differently respectively for $c>2$ and $1<c<2$.

\vskip 0.2cm

\begin{itemize}

\item $c>2$: Consider first the regime $c>2$. In that case the first term in Eq. (\ref{Gr.4}) scales as 
$\sim L^{2-c}$ for large $L$ and hence also be dropped since $c>2$, leaving us with only the 
second term in the thermodynamic limit
\begin{equation}
G(r|L) \approx \frac{y_c}{\langle l\rangle\, (c-1)(c-2)} \frac{1}{r^{c-2}} \quad {\rm for} \quad 1\ll r\ll L
\label{Gr_ucrit.1}
\end{equation}
Finally, in this regime $y=y_c$ and $c>2$, Eq. (\ref{avgl_ucrit.1}) yields $\langle l \rangle \approx
y_c \zeta(c-1)$. Hence, we obtain an $L$ independent correlation function that decays algebraically for large $r$
\begin{equation}
G(r|L) \approx \frac{1}{\zeta(c-1)(c-1)(c-2)}\, \frac{1}{r^{c-2}} \quad {\rm for}\quad r\gg 1 
\label{Gr_ucrit.2}
\end{equation}
Thus, in the case the correlation function is independent of system size $L$ as $L\to \infty$, and behaves as
in a standard critical point with a power law decay of the correlation function, except that the
decay exponent $c-2$ depends continuously on the parameter $c$.
We therefore conclude that for $c>2$, the system does not exhibit FDPO.

\item $1<c<2$: The large $L$ behavior of $G(r|L)$ for $1<c<2$ is drastically different from the $c>2$ case.
In this case, we have to keep both terms in Eq. (\ref{Gr.4}) for large $L$. Furthermore, we see from
Eq. (\ref{avgl_lcrit.1}) that in this case $\langle l \rangle\approx \frac{y_c}{2-c} L^{2-c}$. Substituting
this in Eq. (\ref{Gr.4}) and simplifying we get
\begin{equation}
G(r|L) \approx 1 - \frac{1}{c-1}\, \left(\frac{r}{L}\right)^{2-c}\, .
\label{Gr_lcrit.1}
\end{equation}
In Appendix A, we will provide an alternative derivation of this main result in Eq. (\ref{Gr_lcrit.1}) using IIA.
From Eq. (\ref{Gr_lcrit.1}) we see that
the correlation function, instead of becoming $L$ independent for large $L$ as in the case $c>2$, 
emerges as a function of the scaled distance $u=r/L$ only.
Indeed, the result in Eq. (\ref{Gr_lcrit.1}) is consistent with this scaling picture.
Eq. (\ref{Gr_lcrit.1}) indicates that for large $r$, large $L$ but with the ratio $u=r/L$ fixed, 
the correlation function has a scaling form: $G(r|L)\approx Y(u)$ where the
scaling function $Y(u)$, for $u\ll 1$, behaves as 
\begin{equation}
Y(u) \approx 1 - \frac{1}{c-1}\, u^{2-c}  \quad {\rm for} \quad u\ll 1
\label{Yu.1}
\end{equation}
Thus, the scaling function $Y(u)$ displays a 
cusp singularity of the form in Eq. \ref{Eq_Cusp} with the cusp exponent 
\begin{equation}  
\alpha = (2-c)\, ;   \quad 0<\alpha<1
\end{equation}
A similar relation between the cusp exponent and the exponent characterizing the decay
of the cluster size distribution was found also in the CD model \cite{DBM}.
The variation of the cusp exponent with the value of the TIDSI coupling constant  
should be noted. For $c$ close to 2, the cusp is extremely sharp while as 
$c \rightarrow 1$, the scaling function morphs into $Y(u) \approx 1 - u\, \ln u $. Thus, 
our main conclusion is that the TDSI model, along the critical
line $y=y_c$ and $1<c<2$,
exhibits FDPO with a cusp exponent $2-c$ that varies continuously with $c$, as $c$ varies
between $1$ and $2$.

\end{itemize}

\section{Conclusion}

In our study of the critical line of the TIDSI model, evidence of FDPO 
comes from the occurrence of anomalously large fluctuations, as well as a 
characteristic scaling form of the correlation function. For a range of the parameter $c$, ($1<c<2$), 
the length of the longest cluster $l_{max}$ is of order system size $L$, 
and the distribution of the ratio $l_{max}/L$ approaches an asymptotic 
form as $L \rightarrow \infty$, implying fluctuations of $l_{max}$ are 
anomalous and do not damp out in the thermodynamic limit. Analogous 
fluctuations are expected in other quantities as well, as we will discuss 
below. The other signature of FDPO seen in the model is the cusp 
singularity in the scaled correlation function (Eq. \ref{Eq_Cusp}).

Two points are worth noting. First, within the TIDSI model, the value of the 
cusp exponent $\alpha$ is found to vary continuously with $c$, along a 
portion of the critical line. While it is known that critical exponents 
may vary with parameters along a normal line of critical points in certain 
cases, this is the first example of a similar variation in FDPO. Secondly, 
the TIDSI model represents an equilibrium system, in contrast to the driven, 
nonequilibrium systems in which FDPO was found and studied earlier 
\cite{DB,DBM,MR,DDR,SDR,DPR,KBB,CB}. Thus it is not the fact of 
equilibrium or otherwise that is primarily responsible for FDPO; rather, 
it appears to be the existence of {\it long-ranged interactions}, which 
are manifest in the TIDSI spin Hamiltonian, and can be induced between 
particles through surface fluctuations, in the sliding particle model.

It is interesting to compare the results obtained for the TIDSI model with 
those obtained for a coarse-grained (CD) depth model. The CD model 
corresponds to the extreme adiabatic limit for hard-core particles sliding 
passively on a fluctuating surface \cite{DB,DBM}, and can be interpreted 
as a tied down renewal process on a Brownian bridge, for which analytic 
calculations can be performed \cite {CG1,CG2}. For this model, the 
distribution of $l_{max}$ can be calculated exactly; as for the TIDSI model, it 
is a function of $l_{max}/L$, with multiple mild singularities \cite{CG1}. 
Further, the domain-size distribution also follows a power law in both 
models. Finally, the correlation function is a scaling function of 
$l_{max}/L$ and displays the signature cusp singularity in both CD 
\cite{CG2} and TIDSI models.

The perspective provided by FDPO suggests some natural questions for 
investigation within the TIDSI model. For instance, order parameter 
distributions for the sliding particle and CD models are known to be broad 
in the thermodynamic limit \cite{DBM,KBB} suggesting that a similar result 
should hold for the magnetization in the TIDSI model as well. Further, 
interesting questions arise for the dynamics. The approach to a steady 
state displaying FDPO follows coarsening dynamics, with the correlation 
function following a scaling form as in the steady state, except that $L$ 
is replaced by a characteristic length scale which grows as $\sim t^{1/z}$, where
$z$ is the dynamical exponent. It 
would be interesting to check this within the TIDSI model, and to see 
whether the dynamical exponent $z$ depends on $c$. Finally, the scaled 
two-time autocorrelation function was shown to have a cusp singularity in the 
sliding particle context, and was found analytically in the CD model 
\cite{CB}, suggesting a similar behavior may hold in the TIDSI model as well. 
It would be valuable to investigate and understand these dynamical issues in the
TIDSI model.

Finally, given our results for FDPO within the TIDSI model, the question 
arises whether there is a relationship between FDPO and MOTs in a broader 
context. Indeed, examination of the phases of sliding particles with 
hard-core interactions interacting with a surface reveal an interesting 
scenario. The symmetric Lahiri-Ramaswamy (LR) model, in which the 
particle-surface interactions act synergetically to produce a 
macroscopically large valley, shows a fluctuationless strongly phase 
separated (SPS) state \cite{LBR}. This state is separated from a 
disordered state by a critical line along which the sliding particles are 
passive and do not influence the surface \cite{DB,DBM}; the full phase 
diagram is discussed in \cite{RBDB}. The order parameter shows a 
corresponding 0-1 jump from the disordered to ordered phase, while the 
passive particle problem along the critical line exhibits FDPO. From the 
disordered side, a divergence of the correlation length appears to be 
likely, but has not yet been established. Likewise, a recent study of the 
Light-Heavy (LH) model of particles on a surface \cite{CPCB,CCB} has revealed a 
rich phase diagram with a disordered phase, and several types of ordered 
phases. Interestingly, the separatrix between disordered and ordered 
phases again reduces to a passive scalar problem, except that the driving 
surface follows Kardar Parisi Zhang (KPZ) dynamics in this case, rather 
than the Edwards-Wilkinson driving which operates in the symmetric LR 
model. Thus the state along the critical line is again characterized by FDPO, 
and it would be interesting to check whether there is a MOT in the LH model.

\appendix

\section{Independent Interval Approximation (IIA)}

We have seen in Section 3 that for $y\ge y_c$, the joint $m$-point distribution function of domain sizes
$P(l_1,\, l_2,\, \cdots, l_m)$ becomes factorised in the thermodynamic limit (see Eq. (\ref{IIA.1})). This means that asymptotically
for large $L$, the domains become statistically independent. In other words, the independent interval approximation (IIA)
is actually asymptotically exact. Using IIA, many quantities can be computed analytically~\cite{IIA}, as for the CD model \cite{DBM}. Here we briefly
recall this method and use it to estimate the spin-spin correlation function in our model on the critical line $y=y_c$.
Even though in our problem we have a lattice of finite size $L$, 
if we are interested in distance scales much bigger than the lattice spacing, we can approximate our lattice by
a continuous line. Moreover, we will assume that the line is infinite in the thermodynamic limit.
The line consists of intervals (domains)
separated by the domain walls and we assume that each interval is drawn independently
from a normalized PDF $P(l)$ with a finite first moment $\langle l \rangle= \int_0^{\infty} l\, P(l)\, dl$.
Note that $1/{\langle l\rangle}$ is just the density of domain walls per unit length. Let us also define
$p_n(r)$ as the probability that a segment of length $r$ contains exactly $n$ domain walls. The goal is to estimate
$p_n(r)$ using IIA and then use it to estimate the spin-spin correlation function using the exact identity in
Eq. (\ref{gencorr.1}) namely,
\begin{equation}
G(r)= \sum_{n=0}^{\infty} (-1)^n\, p_n(r)\, ,
\label{gencorr.A1}
\end{equation}

We now outline the derivation of $p_n(r)$ that was worked out in detail in Ref.~\cite{IIA} in a different context.
It is useful to first define the cumulative interval size distribution
\begin{equation}
Q(l)= \int_l^{\infty} P(l')\, dl'\, .
\label{cumQl}
\end{equation}
Thus $dQ/dl= - P(l)$. Consider a segment of total length $r$ with $n$ domain walls.
Hence there are $n+1$ intervals of lengths say $\{l_1,\, l_2,\, \cdots\, l_n,\, l_{n+1}\}$
such that $l_1+l_2+\cdots+l_n+l_{n+1}=r$.
Now, for $n\ge 1$, treating the domains as statistically independent, the probability
$p_n(r)$ can be expressed in terms of $P(l)$ and $Q(l)$ as follows~\cite{IIA}
\begin{equation}
p_n(r)= \frac{1}{\langle l\rangle} \int_0^{\infty} dl_1 \int_{0}^{\infty}dl_2 \cdots \int_{0}^{\infty} dl_{n+1} \, 
Q(l_1)\, P(l_2)\,P(l_3)\cdots P(l_n)\, Q(l_{n+1})\, \delta(l_1+l_2+\cdots +l_n+l_{n+1}-r)\, .
\label{IIA.2}
\end{equation}
The interpretation is straightforward: Given that a domain wall occurs in the interval $r$ (which happens
with probability $1/\langle l \rangle$ per unit length), only the leftmost and the rightmost intervals
are incomplete, explaining the $Q(l)$ at the two ends. In between, $n-1$ intervals are complete, 
each independently with probability density $P(l)$. The presence of the delta function ensures 
that the sum of interval lengths is $r$. The integral can be performed readily in the Laplace space. 
We define ${\tilde p}_n(s)=\int_0^{\infty}p_n(r)\, e^{-s\, r}\, dr$ and 
${\tilde P}(s)= \int_0^{\infty}P(l) e^{-s\, l}\, dl$. Next we take Laplace transform of Eq. (\ref{cumQl}) and
use the relation $Q'(l)=-P(l)$ that gives, after straightforward algebra~\cite{IIA}
\begin{equation}
{\tilde p}_n(s)= \frac{1}{{\langle l \rangle} s^2}\left[1- {\tilde P}(s)\right]^2 \left[{\tilde P}(s)\right]^{n-1}\quad {\rm for}\quad n\ge 1
\label{IIA.3}
\end{equation}
The probability $p_0(r)$ can be estimated from the normalization: $\sum_{n=0}^{\infty} p_n(r)=1$ which gives
$\sum_{n=0}^{\infty} {\tilde p}_n(s)= 1/s$. Using the results for ${\tilde p}_n(s)$ for $n\ge 1$ in 
Eq. (\ref{IIA.3}), we get our desired expression
\begin{equation}
{\tilde p}_0(s)= \frac{1}{{\langle l \rangle} s^2}\,\left[\langle l\rangle s-1+ {\tilde P}(s)\right] 
\label{IIA_p0}
\end{equation}

Let us now focus on the critical line $y=y_c$ with $1<c<2$, where we expect FDPO to manifest.
In this regime, we have from Eq. (\ref{Pl_lcrit.3}), $P(l)\sim y_c/l^c$ for $1\ll l\ll L$. Consequently,
$\langle l\rangle\approx  y_c L^{2-c}/(2-c)$ from Eq. (\ref{avgl_lcrit.1}). Hence, its Laplace transform
${\tilde P}(s)$ has the small $s$ behavior
\begin{equation}
{\tilde P}(s)\approx 1 + y_c\, \Gamma(1-c)\, s^{c-1}+ \cdots
\label{Ps.1}
\end{equation}
Substituting this result in Eq. (\ref{IIA.3}), we get the leading small $s$ behavior for $n\ge 1$
\begin{equation}
{\tilde p}_n(s) \sim \frac{1}{\langle l\rangle}\, \frac{1}{s^{4-2c}}\quad {\rm for\,\, all} \quad n\ge 1\, .
\label{pns.1}
\end{equation}
Inverting the Laplace transform, we then get the following power law tail for $p_n(r)$ for $r\gg 1$ and for any $n\ge 1$
\begin{equation}
p_n(r) \sim \frac{1}{\langle l\rangle}\, \frac{1}{r^{2c-3}}\, .
\label{pnr.1}
\end{equation}
Using $\langle l\rangle\approx y_c L^{2-c}/(2-c)$ for large $L$, we get for $n\ge 1$
\begin{equation}
p_n(r) \sim L^{1-c} \left(\frac{r}{L}\right)^{3-2c} \, .
\label{pnr.2}
\end{equation}
Hence, in the scaling regime with $r$ large, $L$ large, but the ratio $u=r/L$ held fixed, we see from
the prefactor $L^{1-c}$ (recall $c>1$) that all 
$p_n(r)$'s with $n\ge 1$ decay to $0$ as $L\to \infty$. Hence, the $n\ge 1$ terms do not
contribute to the correlation function
$G(r)$ in Eq. (\ref{gencorr.A1}), as we had argued in the main text to obtain Eq. (\ref{corr.2}).

We now turn to $p_0(r)$ and provide an alternative derivation of our result in Eq. (\ref{Gr_lcrit.1}) using
this IIA method. Substituting the small $s$ behavior of ${\tilde P}(s)$ from Eq. (\ref{Ps.1}) into
Eq. (\ref{IIA_p0}) we get the following small $s$ behavior of ${\tilde p}_0(s)$
\begin{equation}
{\tilde p}_0(s) \approx \frac{1}{s} + \frac{y_c\, \Gamma(1-c)}{\langle l\rangle}\, s^{c-3} + \cdots
\label{p0s.1}
\end{equation}
Inverting this Laplace transform and using $\langle l\rangle \approx y_c L^{2-c}/(2-c)$ we get for $1\ll l\ll L$
\begin{equation}
p_0(r)\approx 1- \frac{1}{c-1}\, \left(\frac{r}{L}\right)^{2-c} \, . 
\label{p0r.A1}
\end{equation}
Using $G(r) \approx p_0(r)$ then leads to the result exhibiting FDPO
\begin{equation}
G(r) \approx 1- \frac{1}{c-1}\, \left(\frac{r}{L}\right)^{2-c}\, .
\label{Gr_alt.1}
\end{equation}
This thus provides an alternative derivation of our main result on FDPO (for $y=y_c$ and $1<c<2$),
that was derived by a different method in Section IV.




\subsection*{Acknowledgements}

MB acknowledges useful discussions with C. Godr\`eche. This work was supported by
	a research grant from the Center for Scientific Excellence at the Weizmann Institute of Science. SNM acknowledges the visiting Weston fellowship, and MB and SNM acknowledge the hospitality of 
the Weizmann Institute during the SRITP workshop
``Correlations, Fluctuations and anomalous transport in systems far from
equilibrium" held at the Weizmann Institute in January 2018.



\end{document}